# Positional disorder of Ba in the thermoelectric germanium clathrate $Ba_6Ge_{25}$

V.Petkov[1] and T.Vogt[2]

[1]Department of Physics, Central Michigan University, Mt. Pleasant, MI 48859
[2]Physics Department, Brookhaven National Laboratory, Upton, NY 11973

(March 17, 2003)

**Abstract**

The local structure of $Ba_6Ge_{25}$ has been studied by x-ray diffraction and the atomic pair distribution function technique at 40 K and room temperature. Unambiguous evidence has been found that two out of three types of Ba atoms in $Ba_6Ge_{25}$ move off their positions and become locked in split sites at low temperatures.

Clathrates, in particular $Ba_6Ge_{25}$, have received much attention recently because of their potential as thermoelectric materials [1]. It has been argued that a good thermoelectric material should possess low thermal conductivity and high electric conductivity, i.e. should share the characteristics of both "phonon gases" and "electron crystals" [2]. These two prerequisites are well met with $Ba_6Ge_{25}$ in which electropositive Ba cations are enclosed within oversized cavities of a crystalline framework of covalently bonded Ge anions. The framework is 3D ordered and rigid, and provides a medium for high electrical conductivity of ~ 2000 S/cm [3]. On the other hand, at room temperature the heavy Ba cations rattle inside the framework cavities creating conditions for extensive phonon scattering that leads to low thermal conductivity of ~ 1.2 W/Km. A direct evidence for this rattling has come from diffraction experiments showing quite large values for the atomic displacement parameters of Ba atoms [3-5]. However, the electrical and thermal conductivity of $Ba_6Ge_{25}$ change below room temperature spoiling material's thermoelectric characteristics. The change is due to a structural phase transition taking place between 180 – 215 K [3,5,6].

At room temperature, i.e. above the phase transition, $Ba_6Ge_{25}$ is cubic (S.G. $P4_132$) with four formula units per unit cell [3,5] and a cell parameter of 14.554 Å. Ge atoms are arranged into a three dimensional, non-space filling framework of $Ge_{20}$ dodecahedra. In the structure there are three sites for the Ba atoms each having a different coordination environment. According the atom labeling scheme introduced in [3] the atoms of the Ba(1) site occupy the centers of $Ge_{20}$ polyhedra. The atoms of Ba(2) and Ba(3) sites lie along the channels inside the framework and thus have ample space to rattle. Fragment of the $Ba_6Ge_{25}$ structure is shown in Fig. 1. It has been suggested that upon cooling, Ba(2) and Ba(3) atoms move off their position and become locked in well-separated positions, split sites, inside the channels [5,6]. It is speculated that the locking-in process results in Ba-Ba distances as short as 4.03 Å that are dynamically stabilized by two electrons giving rise to bipolarons. The formation of such polaronic quasiparticles is evoked to explain the observed strong reduction of the carrier mobility at low temperatures [6]. Also, it has been predicted that a shifting of Ba(2) and Ba(3) atoms toward split positions in the framework channels would reduce the density of states at the Fermi level which may account for the observed change in the transport properties at low temperatures [7]. Obviously understanding the properties of $Ba_6Ge_{25}$ requires a detailed knowledge of the changes its atomic structure undergoes with temperature. Here we report results of a structural study addressing this still unsolved issue. We employ the atomic Pair Distribution Function (PDF) technique which has emerged recently as a powerful tool for structural characterization of crystalline materials with intrinsic disorder [8] and nanocrystals [9], including arrays of metal atoms confined within channels of a rigid matrix [10].

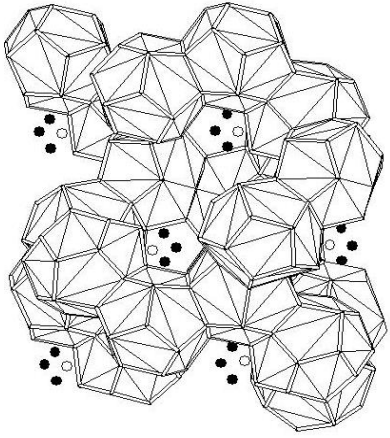

Figure 1. Fragment of the structure of $Ba_6Ge_{24}$ at room temperature illustrating a helical framework of $Ge_{20}$ dodecahedral units with Ba atoms occupying the framework channels. Ba(2) and Ba(3) atoms are shown as open and solid circles, respectively. Structure data are from [3,5].

The frequently used atomic PDF, $G(r)$, is defined as follows:

$$G(r) = 4\pi r[\rho(r) - \rho_o] \quad (1)$$

where $\rho(r)$ and $\rho_o$ are the local and average atomic number densities, respectively and $r$ is the radial distance. $G(r)$ gives the number of atoms in a spherical shell of unit thickness at a distance $r$ from a reference atom. It peaks at characteristic distances separating pairs of atoms and thus reflects the atomic structure. The PDF $G(r)$ is the Fourier transform of the experimentally observable total structure function, $S(Q)$, i.e.

$$G(r) = (2/\pi) \int_{Q=o}^{Q_{max}} Q[S(Q)-1]\sin(Qr)dQ, \quad (2)$$

where $Q$ is the magnitude of the wave vector ($Q=4\pi\sin\theta/\lambda$) and $S(Q)$ is the coherent scattering intensity per atom in electron units [11]. The most important advantage of the PDF technique is that it reflects both the long-range atomic structure, manifested in the Bragg peaks, and the local structural imperfections, manifest in the diffuse components of the diffraction data. In addition, the PDF is a sensitive, structure-dependent quantity that gives directly the relative positions of atoms in materials. This enables the convenient testing and refinement of structural models.

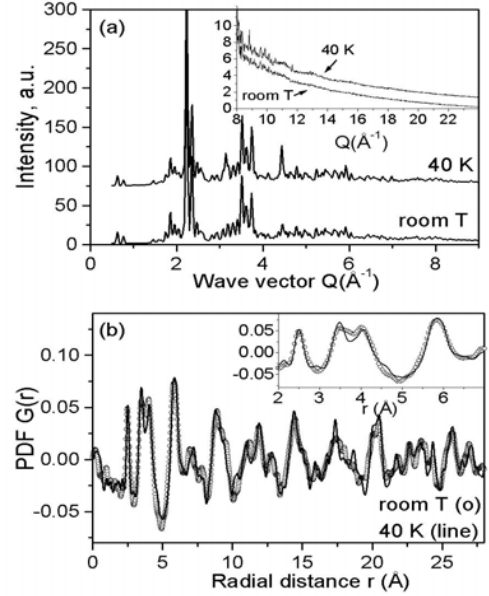

Figure 2. Powder diffraction patterns (a) and the corresponding atomic PDFs (b) of $Ba_6Ge_{24}$ obtained at room temperature and 40 K. Diffraction data are shifted for clarity. A portion of the data is given on an enlarged scale in the insets.

Experimental powder diffraction patterns of $Ba_6Ge_{25}$ obtained at room temperature and 40 K are shown in Fig. 2(a). The experiments were done at the beam line X7A of the National Synchrotron Light Source, Brookhaven National Laboratory with x-rays of energy 29.09 keV ($\lambda$=0.425). Well defined Bragg peaks are seen in both diffraction patterns at low values of Q and, as can be expected, the peaks sharpen when temperature decreases (see the inset in Fig. 2a). In addition, both diffraction patterns show a pronounced, slow oscillating diffuse component which dominates the data at high values of Q (see the inset in Fig. 2a). The component reflects the presence of significant static structural disorder in $Ba_6Ge_{25}$ at both room temperature and 40 K. Information about any kind of atomic disorder is difficult to obtain by usual crystallographic analysis relying on the Bragg scattering only. The problem may be solved by considering the diffraction data in real space in terms of the corresponding atomic PDFs which take both the Bragg and diffuse scattering into account. Experimental PDFs obtained from the diffraction data of Fig. 2(a) are shown in Fig. 2(b).

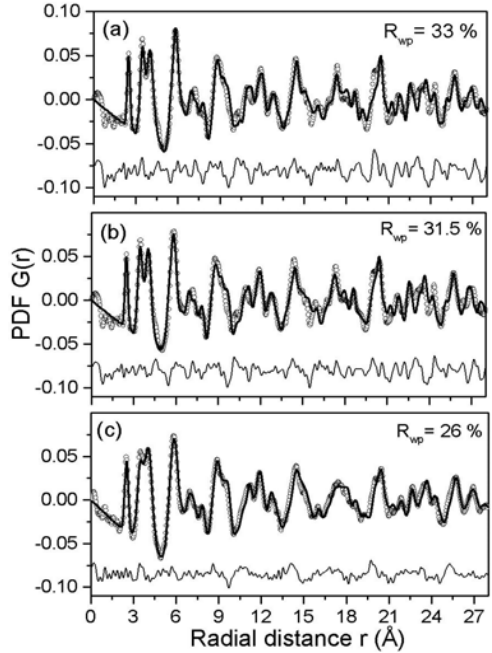

Figure 3. Experimental (dots) and fitted (solid line) PDFs for $Ba_6Ge_{24}$ at room (c) temperature and 40 K (a) and (b). The residual difference between the experimental and model data is shown in the lower part of the three panels as solid line. The corresponding goodness-of-fit indicators, $R_{wp}$ [14], are reported in each of the panels.

Details of x-ray data processing and PDF derivation are described elsewhere [11,12].

As can be seen in Fig. 2(b) both PDFs of $Ba_6Ge_{25}$ show well defined peaks extending to high real space distances as it should be with a crystalline material. The first peak is positioned at ~ 2.5 Å and corresponds to the first neighbor Ge-Ge atoms sitting on the vertices of the $Ge_{20}$ units shown in Fig. 1. The peak does not change with temperature (see the inset in Fig. 2b) which shows that the immediate atomic ordering and connectivity of the framework of Ge atoms are preserved upon going through the phase transition. The second PDF peak positioned at distances from 3 to 5 Å, however, changes its shape with temperature. The change (see the inset in Fig. 2b) involves inverting the ratio of the two components of the peak with the lower-r one becoming more intense at low temperatures. Since the first neighbor Ba-Ge and Ba-Ba distances in $Ba_6Ge_{25}$ occur at 3.5 Å and 4.7 Å, respectively, the observed change may only come from a *static shift* of Ba atoms. This is a clear evidence that some Ba atoms in $Ba_6Ge_{25}$ move off their positions upon decreasing temperature. The shift of Ba atoms affects more distant PDF peaks as well with the most noticeable change is observed as an extra sharpening of the PDF peak at ~ 17.5 Å (see Fig. 2b).

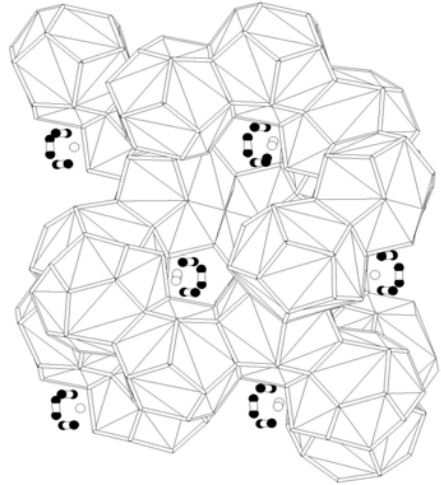

Figure 4. Fragment of the structure of $Ba_6Ge_{24}$ at 40 K as deducted by the present PDF study. The structure features a helical array of $Ge_{20}$ dodecahedral units with Ba atoms occupying split positions inside the array's channels. Ba(2) and Ba(3) atoms are shown as split open and solid circles, respectively.

To reveal the way Ba atoms rearrange with temperatures we fit the experimental PDF data with structure models based on reported structure data for $Ba_6Ge_{25}$ [3,5]. The fit was done with the help of the program PDFFIT [13] and involved a single unit cell of $Ba_6Ge_{25}$ having 124 atoms. A comparison between the experimental and model calculated PDFs is shown in Fig. 3. As can be seen in Fig. 3(c) a model based on the reported room temperature structure of $Ba_6Ge_{25}$ is capable of reproducing the experimental PDF data in finest detail. Note that this very good agreement was achieved without further refining the positions of Ge and Ba atoms in the cubic unit cell; only the corresponding overall thermal parameters were adjusted slightly. Those parameters refined to $u(Ba1) = 0.011$ Å$^2$, $u(Ba2) = 0.028$ Å$^2$, $u(Ba3) = 0.070$ Å$^2$ and $u(Ge) = 0.016$ Å$^2$. Thus the present PDF study confirms the presence of a considerable positional disorder on Ba(2) and Ba(3) sites at room temperature. The most noticeable changes in the experimental PDF data observed at low temperatures, that at ~ 4 Å and 17.5 Å, were possible to be reproduced only when the constraints of S.G. $P4_132$ were relaxed and Ba(2) and Ba(3) atoms were let to move off their sites and occupy split positions in the channels of Ge framework (see Fig. 3a). The moves amounted to ~ 0.6 Å and resulted in a change of the position of Ba(2) atoms from (0.625,0.875,0.125) [3] to (0.3718,0.3718,0.3718) and that of Ba(3) atoms from (0.1905,0.4405,0.125) [3] to (0.1889,0.4370, 0.1054). At the same time, the atomic thermal parameters refined to the reasonable for low temperatures values of $u(Ba1) = 0.004$ Å$^2$, $u(Ba2) = 0.009$ Å$^2$, $u(Ba3) = 0.008$ Å$^2$ and $u(Ge) = 0.011$ Å$^2$. Various distributions of Ba(2) and Ba(3) atoms over the split positions were tested but none was found to give a significantly better fit to the experimental PDF data than the others. The result indicates that, to a good first approximation, Ba(2) and Ba(2) atoms distribute randomly on the split sites in the channels of the Ge framework as shown in Fig. 4.

From a structural point of view the observed shifting of Ba atoms toward one of the split positions may be viewed as a lowering of the local symmetry of Ba(2) and Ba(3) sites from S.G. $P4_132$ to S.G. $P2_13$ as discussed in Ref. [7]. In line with previous structural studies [4,6] the present one did not find any tendency of Ba(1) atoms to move off the centers of Ge dodecahedra upon going through the phase transition.

Although the model featuring split Ba(2) and Ba(3) positions managed to reproduce the most noticeable changes in the experimental PDF at low temperature, the level of agreement between the model and experimental PDF data appears somewhat worse than that achieved with the room temperature data, as a comparison between the results presented in Fig. 3(a) and Fig. 3(c) shows. We attempted to refine the low temperature structure model further by changing the amplitude of splitting of Ba(2) and Ba(3) sites. A model PDF obtained by averaging 10 structure models each with a slightly different amplitude of splitting of Ba(2) and Ba(3) sites, varying from 0.5 Å to 1.2 Å, is shown in Fig. 3(b). The agreement with the experimental data is improved which shows that different Ba(2) and Ba(3) atoms are likely to move off their positions with slightly different amplitudes when temperature is reduced. Thus not only the distribution of Ba atoms over the split sites but the amplitude of splitting of those sites too seem to show some degree of randomness. This may well explain the absence of any superstructure and the apparent enhancement of the diffuse component of diffraction data (see the inset in Fig. 2a) at low temperatures.

Attempts to further improve the agreement with the experimental data by not only switching Ba(2) and Ba(3) atoms between varios split sites with local symmetry S.G. $P2_13$ but also refining the positions of Ge atoms within the constraints of S.G. $P4_132$ failed. The likely reason is that Ge framework too relaxes at low temperatures and may not be anymore described in terms of S.G. $P4_132$. We did not pursue the temperature driven changes in the Ge framework since this required extensive modeling going beyond the scope of the present study.

In summary, the present PDF study produced an unambiguous evidence that Ba(2) and Ba(3) atoms in the clathrate $Ba_6Ge_{25}$ shift toward split positions at low temperatures and yielded the coordinates of those positions. Also, it found that the splitting of Ba(2) and Ba(3) sites and the distribution of Ba atoms over them do not follow a particular pattern. The new structural information published here could be used in more precise electronic structure calculations of this clathrate material revealing new insight into its thermoelectric properties.

Thanks are due to M. Schmidt, P.G. Radaelli, N. Hur and S. W. Cheong for providing the sample and fruitful discussions. This work was partially supported by NSF through Grant CHE-0211029. NSLS is supported by the US Department of Energy through DE-AC02-98-CH0886.

[1] B. C. Sales, Science 295 (2002) 1248.
[2] G.S. Nolas, G.A. Slack, D.T. Morelli, T.M. Tritt and A.C. Erlich, J. Appl. Phys. 79 (1996) 4002; G.A. Slack, Solid State Physics 34 (1979) 1.
[3] S-J. Kim, S. Hu, C. Uher, T. Hogan, B. Huang, J.D. Corbett and M.G. Kanatzidis, J. Solid State Chem. 153 (2000) 321.
[4] M. Schmidt and P. Radaelli, *private communication*.
[5] F. Fukuoka, K. Ieai, S. Yamanaka, H. Abe, K. Yoza and L. Haming, J. Solid State Chem. 151 (2000) 117.

[6] S. Paschen, V.H. Tran, M. Baenitz, W. Carrillo-Cabrerra, Yu. Grin and F. Steglich, Phys. Rev. B 65 (2002) 134435.
[7] I. Zerec, A. Yaresko, P. Thalmeier and Yu. Grin, Phys. Rev. B. 66 (2002) 045115.
[8] V.Petkov, I-K. Jeong, J.S. Chung, M.F. Thorpe and S.Kycia and S.J.L. Billinge Phys. Rev. Lett. 83 (1999) 4089; S.J.L. Billlinge, Th. Proffen, V. Petkov, J.L. Sarrao and S. Kycia, Phys. Rev B 62 (2000) 1203
[9] V. Petkov, E. Bozin, S.J.L. Billinge, T. Vogt, P. Trikalitis and M. Kanatzidis, J. Am. Chem. Soc. 124 (2002) 10157.
[10] V. Petkov, S.J.L. Billinge, T .Vogt, A.Ichimura and J.L. Dye, Phys. Rev. Lett. 89 (2002) 075502.
[11] H.P. Klug and L.E. Alexander, X-ray diffraction procedures for polycrystalline materials, (Willey: New York, 1974).
[12] V. Petkov, J. Appl. Cryst. 22 (1989) 387.
[13] Th. Proffen and S.J.L. Billinge, J. Appl. Cryst. 32 (1999) 572.
[14] Agreement factors reported here may appear high when compared to the agreement factors usually reported with conventional crystallographic techniques. This does not indicate an inferior structure refinement but merely reflects the fact that, the atomic PDF being fit differs from the XRD pattern typically fit in a Rietveld refinement and is a quantity much more sensitive to local ordering in materials. As a result, $R_{wp}$'s greater than 15-20 % are common with PDF refinements of well crystallized materials.